# Evidence of itinerant holes for long-range magnetic order in tungsten diselenide semiconductor with vanadium dopants

*Bumsub Song*[†,‡,○], *Seok Joon Yun*[‡,○], *Jinbao Jiang*[†,‡], *Kory Beach*[§], *Wooseon Choi*[†], *Young-Min Kim*[†,‡], *Humberto Terrones*[§], *Young Jae Song*[‡,∥,⊥,#], *Dinh Loc Duong*[†,‡,\*], *and Young Hee Lee*[†,‡,⊥,\*]

[†]Department of Energy Science, Sungkyunkwan University, Suwon 16419, Republic of Korea.

[‡]Center for Integrated Nanostructure Physics (CINAP), Institute for Basic Science (IBS), Suwon 16419, Republic of Korea

[§]Department of Physics, Applied Physics, and Astronomy, Rensselaer Polytechnic Institute, Troy, New York 12180, USA

[∥]Sungkyunkwan Advanced Institute of Nanotechnology (SAINT), Sungkyunkwan University, Suwon 16419, Republic of Korea

[⊥]Department of Physics, Sungkyunkwan University, Suwon 16419, Republic of Korea

[#]Department of Nano Engineering, Sungkyunkwan University, Suwon 16419, Republic of Korea

One primary concern in diluted magnetic semiconductors (DMSs) is how to establish a long-range magnetic order with a low magnetic doping concentration to maintain the gate tunability of the host semiconductor, as well as to increase Curie temperature. Two-dimensional van der Waals semiconductors have been recently investigated to demonstrate the magnetic order in DMSs; however, a comprehensive understanding of the mechanism responsible for the gate-tunable long-range magnetic order in DMSs has not been achieved yet. Here, we introduce a monolayer tungsten diselenide ($WSe_2$) semiconductor with V dopants to demonstrate the long-range magnetic order through itinerant spin-polarized holes. The V atoms are sparsely located in the host lattice by substituting W atoms, which is confirmed by scanning tunneling microscopy and high-resolution transmission electron microscopy. The V impurity states and the valence band edge states are overlapped, which is congruent with density functional theory calculations. The field-effect transistor characteristics reveal the itinerant holes within the hybridized band; this clearly resembles the Zener model. Our study gives an insight into the mechanism of the long-range magnetic order in V-doped $WSe_2$, which can also be used for other magnetically doped semiconducting transition metal dichalcogenides.

A wide range of magnetic semiconductors with tailored spin-polarized carriers has attracted considerable attention for the realization of gate-tunable magnetic functionalities and low power consumption.[1–3] However, applications of most of the intrinsic magnetic semiconductors are hindered by their low Curie temperatures ($T_C$).[3] Semiconductors doped with magnetic elements, called diluted magnetic semiconductors (DMSs), are considered as alternative materials that can be used for achieving high $T_C$ values.[4–7] Recently, two-dimensional (2D) van der Waals (vdW) materials, particularly transition-metal dichalcogenides (TMDs) with magnetic dopants, have been proposed as a new class of DMSs.[8–12] Gate-tunable magnetic domains have been introduced in a semiconducting WSe$_2$ monolayer by injecting V magnetic dopants through chemical vapor deposition (CVD).[10,12,13] Magnetic signatures have also been observed in a Re-doped MoSe$_2$[9] and V-doped MoTe$_2$.[11] The *pd–d* hybridization in the 2D TMDs achieved by the incorporation of *d*-orbitals of both transition metals in the TMDs and magnetic dopants can be more beneficial for a high $T_C$ and may overcome the limitation of *sp–d* hybridization in III–V DMSs.[5,6] However, although the use of 2D TMDs with magnetic dopants can be a promising route to increase $T_C$, the mechanism of the generation of magnetic order in 2D DMSs is not well understood.

Two main mechanisms in DMSs have been proposed: i) direct interactions between dopants through their directly overlapped wavefunctions, which is referred to as impurity band model,[14–16] and ii) indirect interactions through itinerant carriers, which is referred to as Ruderman–Kittel–Kasuya–Yosida (RKKY) or Zener model.[17–22] The former requires a short distance between dopants and is dominant when the Fermi level is pinned inside the localized impurity band. In contrast, the latter involves long-range interactions between distant dopants and occurs upon strong hybridization between the dopant and host orbitals that leaves the Fermi level inside the overlapped bands. The magnetic exchange interaction between the dopants can be easily modulated by itinerant carriers, thereby enabling gate-tunable magnetic properties.

Here, we report a clear evidence of the prevailing Zener model in a V-doped WSe$_2$ monolayer by the presence of the Fermi level inside the *pd–d* hybridized band, thereby leading to a long-range magnetic order. V atoms are incorporated into the WSe$_2$ lattice by substituting W atoms and form impurity states near the band edges. In particular, a strong hybridization occurs between the impurity states and the host valence band, introducing itinerant holes to the host lattice that facilitate the long-range magnetic order through RKKY interactions between the V dopants.

Bare and V-doped WSe$_2$ monolayers were grown on SiO$_2$/Si substrates by CVD and transferred onto highly oriented pyrolytic graphite (HOPG) by using the water-assisted method.[23] After a mild annealing at ~200 °C for 2 h in an ultrahigh vacuum chamber with a base pressure of ~2.0 × 10$^{-11}$ Torr, scanning tunneling microscopy/spectroscopy (STM/S) measurements were performed at 78 K (See the Supporting Information for further details). **Figure 1**a shows an STM image of the bare WSe$_2$ monolayer on HOPG at a sample bias voltage ($V_s$) of -1.5 V. Sub-nanometer point depressions, denoted as a type-A defect, were observed as a majority defect. The close-up STM image of a type-A defect shows that the depression corresponds to the Se site of the WSe$_2$ lattice (inset). The filled-state morphology of the type-A defect varied with $V_s$ from a dot shape at $V_s$ = -1.5 V (**Figure 1**a) to a triangular shape at $V_s$ = -1.7 V, and the empty-state morphology at $V_s$ = 0.8 V still showed a depression (Figure S1, Supporting Information). The depression in both filled- and empty-state STM images indicates that this defect was a genuine pit in the topograph. Then, annular dark-field scanning transmission electron microscopy (ADF-STEM) measurements of the bare WSe$_2$ was carried out to further examine the atomic structure of the type-A defect (**Figure 1**b). A Se monovacancy was identified as a majority defect (Figure S2, Supporting Information). Therefore, we assign the type-A defect as a single Se vacancy ($Se_{vac}$) (**Figure 1**c). Notably, this type of defect has recently been proposed as a chalcogen vacancy passivated by oxygen.[24–26] However, in this study, we could not identify it further owing to the lack of experimental resolution.

**Figure 1**d demonstrates an STM image of the 0.3%-V-doped $WSe_2$ monolayer on HOPG at $V_s$ = -1.5 V. Similar to the bare $WSe_2$, $Se_{vac}$ was also observed in the V-doped $WSe_2$ monolayer. More importantly, another type of defects, denoted as a type-B defect, was clearly manifested as a bright protrusion. Each protrusion spread out to several adjacent Se atoms (~2 nm). To identify the atomic registry of the type-B defect, we analyzed the 0.3%-V-doped $WSe_2$ by STEM as shown in **Figure 1**e. The V dopants are identified as substitutions of W atoms ($V_{Ws}$). The V doping concentration determined by a statistical analysis (Figure S3, Supporting Information) was $(3.00 \pm 2.64) \times 10^{12}$ cm$^{-2}$, which is consistent with the value determined by STM (i.e., $(1.81 \pm 0.530) \times 10^{12}$ cm$^{-2}$ (Figure S4, Supporting Information). Thus, we assign the type-B defects observed in the STM measurement as $V_{Ws}$ (**Figure 1**f). Notably, the $V_{Ws}$ defects were randomly distributed with the nearest V–V atoms sparsely distributed with a separation of ~4 nm. A smaller separation was rarely observed in our 0.3%-V-doped $WSe_2$ monolayer (Figure S4, Supporting Information).

To unambiguously identify the nature of $V_{Ws}$ and its local spectroscopic characteristics, we performed STS measurements at several $V_{Ws}$ sites. **Figure 2**a shows typical d$I$/d$V$ spectra acquired directly on (red curve), near (orange curve), and far from (black curve) the $V_{Ws}$ site (inset). The d$I$/d$V$ spectrum at the $V_{Ws}$ site was markedly distinct from that of the defect-free region. Apparently, three pronounced peaks (denoted as $IS_1$, $IS_2$, and $IS_3$) were observed within the bandgap. $IS_1$ was observed at -0.75 eV, merged with the valence band edge, whereas $IS_2$ and $IS_3$ were observed at 1.17 and 1.62 eV, respectively, just below the conduction band edge. In contrast, no such peculiar peaks were observed for the location ~2 nm away from the $V_{W_s}$ site, but the valence band tails were still observed, compared to the defect-free region. This implies persistent impurity states far from the $V_{W_s}$ sites as shown in **Figure 1**d. The upshift of the d$I$/d$V$ spectrum of the $V_{W_s}$ site with respect to that of the defect-free region indicates that $V_{W_s}$ is an acceptor. The statistical analysis of the defect-free regions of the bare and V-doped WSe$_2$ monolayers reveals that the increase in V concentration led to a more p-type WSe$_2$ while it rarely affects the bandgap size of approximately 2.2 eV, consistent with the previous reports[27–29] (Figure S5, Supporting Information).

To determine the origin of these impurity bands, we performed density functional theory (DFT) calculations including the spin–orbit coupling (SOC) (See the Supporting Information for further details). **Figure 2**b depicts the calculated local density of states (LDOS) of the V-doped WSe$_2$ at different positions. The peaks in LDOS were found at the $V_{W_s}$ site (red curve), congruent with the experimental observation (**Figure 2**a). The discrepancy in Fermi level positions between the experiment and theory is attributed to the unintentional charge transfer from the HOPG substrate, which pinned the observed Fermi level position above IS$_1$. This agreement between the experiment and theory supports our claim that the type-B defects prevailing in the V-doped WSe$_2$ are $V_{W_s}$ defects (See other possible candidates of V dopant structures in Figure S6, Supporting Information).

Further, we performed sequential bias-dependent STM imaging at the same position to determine the spatial extent of the electronic perturbation of $V_{W_s}$ (**Figure 2**c). At $V_s$ = -1.3 V, three bright protrusions were observed, originating from the enhanced tunneling owing to the additional density of states of IS$_1$. At $V_s$ = 0.6 V, a broad and weak depression was observed for each $V_{W_s}$. Because $V_s$ within the bandgap led to a reduced tip-to-sample distance owing to the lack of available states for tunneling, a clear atomic resolution could be obtained, which enabled the imaging of all hidden Se$_{vac}$ defects. The sudden change in contrast between the filled- and empty-state STM images suggests that the observed morphology is largely affected by the electronic structure of the acceptor-type dopant and related band-bending,[30,31] rather than that it is a real topograph. A weak protrusion was observed at V$_{Ws}$ sites at $V_s$ = 1.0 V, which corresponds to the impurity state of IS$_2$. Several Se$_{vac}$ defects were still visible at this $V_s$. A deep depression in each V$_{Ws}$ was visible at $V_s$ = 1.8 V, which is attributed to the lower density of states of V$_{Ws}$ than that of the defect-free region above IS$_3$.

To investigate a detailed band structure of V-doped WSe$_2$, we calculated the projected density of states (PDOS) of the *d*- and *p*-orbitals of the corresponding V/W and Se atoms in conjunction with the spin density of states, respectively (**Figure 3**). The Fermi level was located near the valence band edge that is significantly contributed from the *d*-orbitals at the V site. The *d*-orbitals of W and *p*-orbitals of Se near the V site yielded similar density of states (**Figure 3**a). More importantly, the density of states of the *d*-orbitals of W atoms far from the V site were still found at the Fermi level, whereas the density of states of *p*-orbitals of Se atoms far from V sites was negligible at the Fermi level. This indicates a strong *d*(W)–*d*(V) hybridization at the V site and even far from the V site, which mediates the long-range hole–hole interaction.

In addition, we investigated the spin distribution of magnetization along the c-axis (**Figure 3**b and 3c). The side-view spin-density map demonstrates the inhomogeneous up-spins (red) and down-spins (blue) near the V site. Ferromagnetism was manifested with a magnetic moment of 1.43 $\mu_B$ at a V doping concentration of 1.5%. Notably, the spin-up holes were spread out through the W sites far from the V site, implying delocalized spin-polarized carriers in the V-doped WSe$_2$. The spin density from W and Se planes containing W, V, and Se atoms, respectively, shows that the six W atoms near the V site exhibited spin-down electrons, opposite to the spin-up electrons at the V site (**Figure 3**c). However, spin-up states were persistent at the other W sites far from the V site, as well as at the V site. This indicates manifested long-range magnetic order in the diluted V-doped WSe$_2$, supporting the RKKY model. Meanwhile, the spin density at the Se plane revealed negligible spin states at Se sites far from the V site.

Although the density of states observed in the STS is consistent with the DFT results, a discrepancy in Fermi level position was observed (**Figure 2**), which originated from the charge transfer from the HOPG substrate to the V-doped WSe$_2$. To resolve this, we measured the field-effect-transistor characteristics of the bare and 0.3%-V-doped WSe$_2$ structures on SiO$_2$/Si substrates (**Figure 4**). The bare WSe$_2$ exhibited p-type semiconducting characteristics with a threshold of approximately -15 V and on/off ratio of $10^6$ (**Figure 4**a). The V-doped WSe$_2$ exhibited a stronger p-type characteristic with a threshold voltage of 70 V and similar on/off ratio (**Figure 4**b). Remarkably, the d$I$/d$V_g$ characteristic of the V-doped WSe$_2$ (transconductance curve) had a dome shape related to the V impurity states near the valence band edge, which is different from the characteristic of the bare WSe$_2$. This is consistent with the valence band structure from the d$I$/d$V$ spectrum acquired at the V$_{Ws}$ site (**Figure 2**a), which resembles the Zener-type valence band picture for DMS systems[1,32] as illustrated in **Figure 4**c. The Fermi level is located within the $p$(Se)$d$(W)–$d$(V) hybridized band that introduces holes to the host, which implies that the magnetic ordering of the V dopants and adjacent W atoms can be robustly modulated by a gate bias.[10,12]

In summary, the structural and electronic properties of the CVD-grown V-doped WSe$_2$ were analyzed by employing STM/S and STEM. Two prominent defect types were identified: i) single Se-vacancy (Se$_{vac}$) and ii) V dopant substituting W (V$_{Ws}$). Our STEM measurements and DFT calculations supported the claim that the V dopants substituted W atoms, which was responsible for the emergence of the hybridized band between the V impurity states and the valence band edge. Moreover, transport characteristics and DFT calculation further revealed itinerant holes that originated from the V dopants were spin-polarized and thus mediated the long-range magnetic order via RKKY interaction. The results of our study are valuable for gaining a better understanding of the mechanism of magnetic order in 2D TMDs such as the recently reported room-temperature ferromagnetism with a gate tunability in a V-doped $WSe_2$.[10]

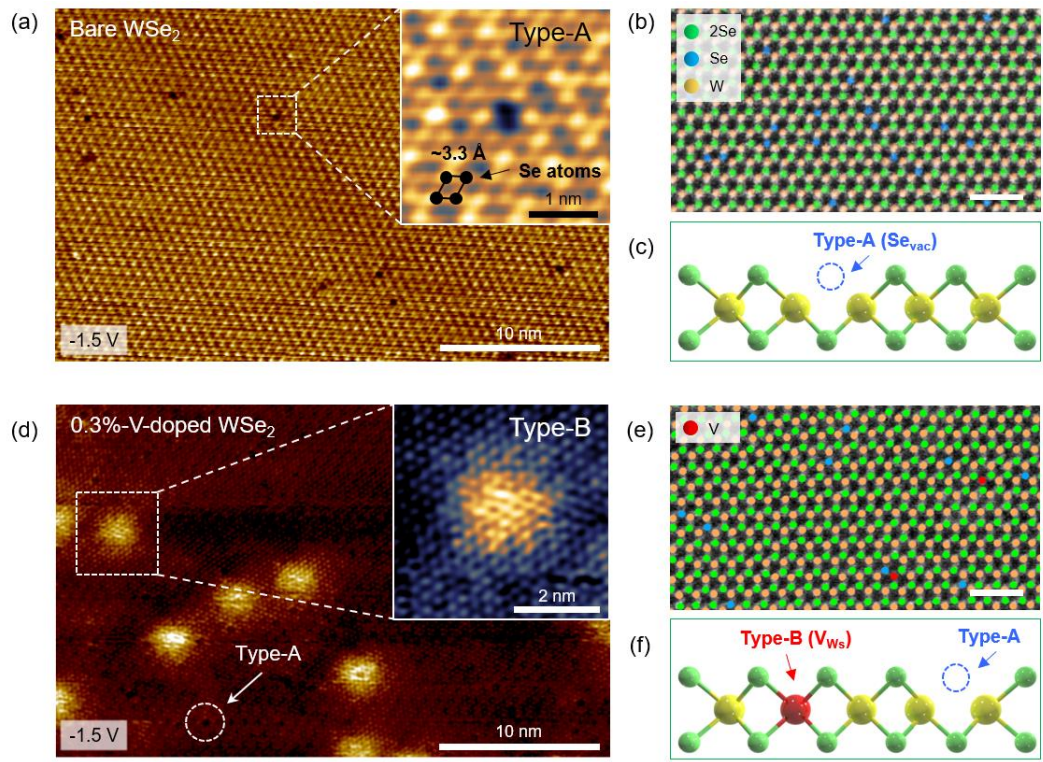

**Figure 1.** Structural properties of the bare and V-doped $WSe_2$. (a) Atomically resolved STM image of the bare $WSe_2$ monolayer. (inset) $Se_{vac}$ assigned as a type-A defect. (b) Atom site mapping for ADF-STEM image of the bare $WSe_2$. (c) Schematic of a single $Se_{vac}$. (d) Atomically resolved STM image of the 0.3%-V-doped $WSe_2$. (inset) $V_{Ws}$ assigned as a type-B defect. A type-A defect ($Se_{vac}$) is outlined by a circle. (e) Atom site mapping for ADF-STEM image of the V-doped $WSe_2$. (f) Schematic of $V_{Ws}$. The scale bars in (b) and (e) represent 1 nm.

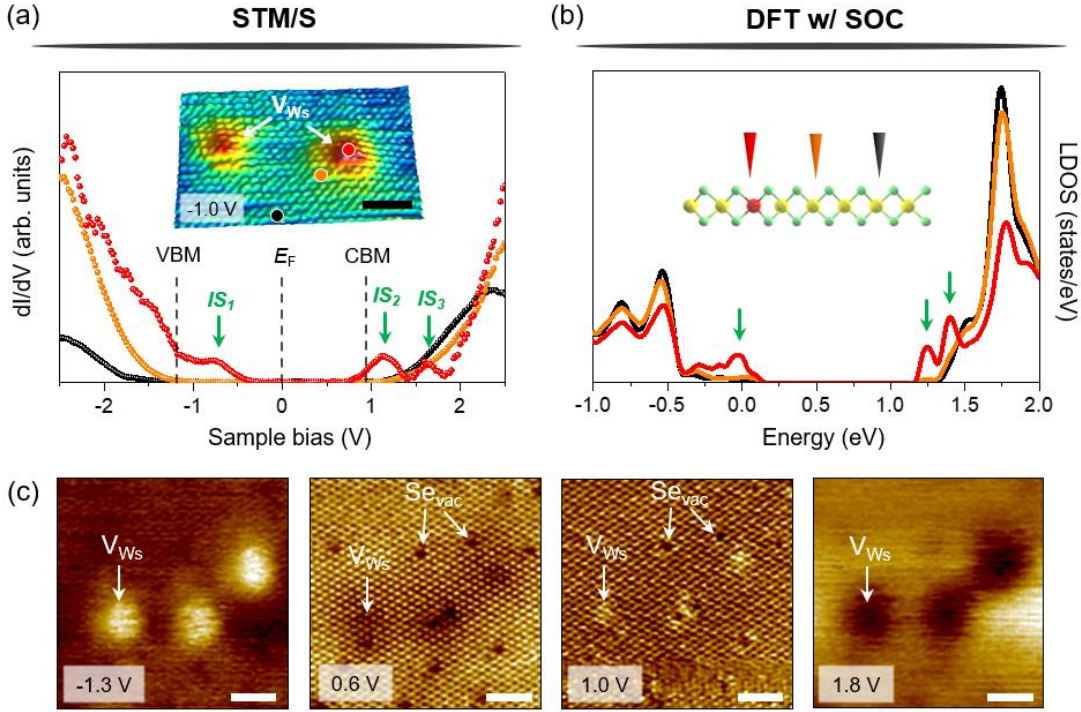

**Figure 2.** STM/S measurements on $V_{W_S}$. (a) Typical d$I$/d$V$ spectra for locations of (red curve), near (orange curve), and far from (black curve) the $V_{W_S}$ site (inset). The impurity states (IS$_1$, IS$_2$, and IS$_3$) that originated from $V_{W_S}$ are marked with arrows. The Fermi level ($E_F$) is outlined at $V_s$ = 0 V. The valence band maximum (VBM) and conduction band minimum (CBM) in the spectrum of the defect-free region are marked. (b) Calculated LDOS of the V-doped WSe$_2$. The states corresponding to the V impurity states in a are marked with arrows. (inset) Structure model of $V_{W_S}$. The atomic positions for each curve are marked. (c) Sequential bias-dependent STM images of $V_{W_S}$ with adjacent Se$_{vac}$ defects. $V_{W_S}$ and Se$_{vac}$ are marked with arrows. The scale bars in (a) and (c) represent 2 nm. Setpoint conditions in (a), $V_s$ = 2.0 V, $I$ = 100 pA.

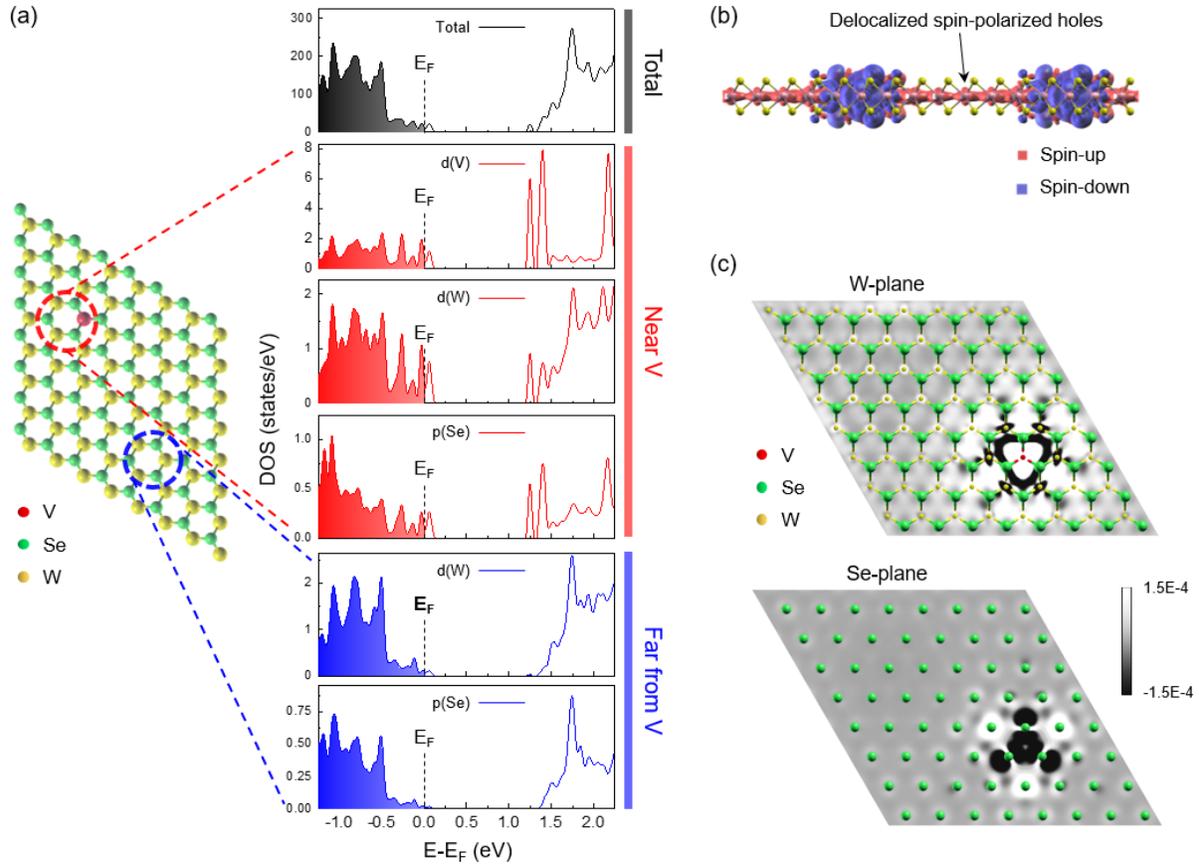

**Figure 3.** Band structure and delocalized spin carriers of the V-doped $WSe_2$. (a) Calculated total density of states and PDOS of the *d*- and *p*-orbitals of the corresponding V/W and Se atoms in the atomic model. (b) Iso-surface of the spin density in real space. The spin down (up) density is presented in blue (red). The isosurface value is 0.00007 $\mu_B/Å^3$. (c) Cross-section spin density map in (b). Positive (negative) values represent spin up (down).

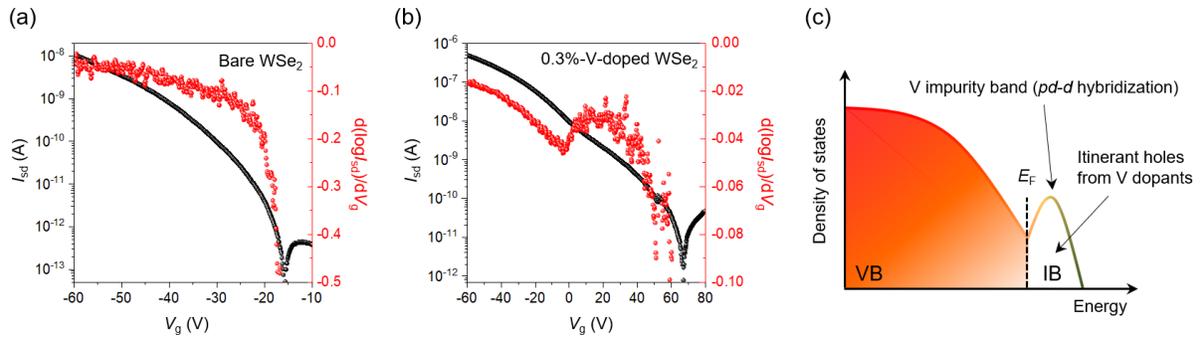

**Figure 4.** Comparison of the transport characteristics of the bare and 0.3%-V-doped WSe$_2$. Transport curves of the (a) bare and (b) 0.3%-V-doped WSe$_2$. 1 V source-drain bias ($V_{sd}$) was applied. (c) Schematic of the valence band structure of the V-doped WSe$_2$.

ASSOCIATED CONTENT

**Supporting Information**.

The following files are available free of charge.

Details on experimental and theoretical methods, STM/S investigation of $Se_{vac}$, density estimation of $Se_{vac}$ and $V_{Ws}$, STS spectra comparing samples with different doping concentration and DFT calculation of other possible structures of V dopants. (PDF)

AUTHOR INFORMATION

**Corresponding Author**

*E-mail (D.L.D.): ddloc@skku.edu.

*E-mail (Y.H.L.): leeyoung@skku.edu.

**Author Contributions**

°B.S. and S.J.Y. contributed equally.

**Notes**

The authors declare no competing financial interest.

ACKNOWLEDGMENT

This work was supported by the Institute for Basic Science (IBS-R011-D1). K.B. and H.T. are grateful to the National Science Foundation (EFRI-1433311). Supercomputer time was provided by the Center for Computational Innovations (CCI) at Rensselaer Polytechnic Institute and the Extreme Science and Engineering Discovery Environment (XSEDE, project TG-DMR17008).

Y.-M.K. acknowledges partial financial support by Creative Materials Discovery Program (NRF-2015M3D1A1070672) through the NRF grant.